\documentclass{article}

\usepackage[preprint]{neurips_2020}

\usepackage[utf8]{inputenc} 
\usepackage[T1]{fontenc}    
\usepackage{hyperref}       
\usepackage{url}            
\usepackage{booktabs}       
\usepackage{amsfonts}       
\usepackage{nicefrac}       
\usepackage{microtype}      

\usepackage{multicol}
\usepackage{algorithm}
\usepackage{amsmath}
\usepackage{graphicx}
\usepackage{subcaption}
\usepackage{float}
\usepackage{algpseudocode} 
\usepackage{xspace}
\usepackage{xcolor}
\usepackage[numbers]{natbib}

\usepackage{mathtools}
\usepackage{bbm}
\usepackage{xspace}
\usepackage{multirow}
\usepackage{diagbox}
\usepackage{ctable}
\usepackage{xcolor}
\usepackage{caption}
\usepackage{subcaption}
\usepackage{amsfonts}
\usepackage{wrapfig}

\DeclareRobustCommand\onedot{\futurelet\@let@token\@onedot}
\def\onedot{\ifx\@let@token.\else.\null\fi\xspace}

\usepackage{subcaption}
\newcommand{\BibTeX}{\rm B\kern-.05em{\sc i\kern-.025em b}\kern-.08em\TeX}

\title{Satellite Navigation and Coordination with Limited Information Sharing}

\author{%
    Sydney Dolan\textsuperscript{\rm 1}, 
    Siddharth Nayak\textsuperscript{\rm 1},
    \textbf{Hamsa Balakrishnan}\textsuperscript{\rm 1}\\
    \textsuperscript{\rm 1} Massachusetts Institute of Technology\\
    \texttt{\{sydneyd, sidnayak, hamsa\}@mit.edu}\
}

\begin{document}

\maketitle
\begin{abstract}
      We explore space traffic management as an application of collision-free navigation in multi-agent systems where vehicles have limited observation and communication ranges. We investigate the effectiveness of transferring a collision avoidance multi-agent reinforcement (MARL) model trained on a ground environment to a space one. We demonstrate that the transfer learning model outperforms a model that is trained directly on the space environment. Furthermore, we find that our approach works well even when we consider the perturbations to satellite dynamics caused by the Earth's oblateness. Finally, we show how our methods can be used to evaluate the benefits of information-sharing between satellite operators in order to improve coordination. 
\end{abstract}

\section {Introduction}
   There are an estimated 8,800 satellites and over a million pieces of debris in the sky today \citep{space_debris_office_2022}; by 2030, there will be an estimated 150,000 active satellites in space \citep{ocallaghan_2022}. The sheer number of objects in orbit and the resulting potential collisions will likely make current approaches untenable, and make autonomous decision-making an essential characteristic of space traffic management (STM) in the future \citep{hobbs_collins_feron_2020}.

    Multi-agent reinforcement learning (MARL) has yielded promising results in several settings \citep{AlphaStar,Dota2}, including for navigation and collision avoidance problems \citep{MADDPG,MAPPO}. Transfer learning has achieved extensive success by leveraging prior knowledge of past learned policies of relevant tasks \citep{transfer_learn,GPG,Muandet_2017}.  Inspired by this, we investigate the effectiveness of transferring a ground-based collision avoidance MARL model to space traffic management, or more specifically, for the collision-free navigation of satellites in orbit. We leverage our recent work on a graph neural network (GNN) based architecture for MARL, called InforMARL \citep{informarl_icml}. We demonstrate that transfer learning from the ground to the space environment is remarkably effective: it achieves better sample complexity and slightly higher rewards than when directly training a model in the space environment. This is despite the two environments being quite different in terms of the underlying dynamics that govern them. We then consider a more refined abstraction of the space environment that accounts for perturbations in gravitational disturbances due to the Earth's oblateness and find that transfer learning is still effective.   

    Finally, we study the role that information-sharing plays in satellite operator decision-making. Operators are hesitant to share information about their satellites for a number of reasons, including  proprietary and security concerns \citep{rendleman_mountin_2015}. While third-party screening services may scan for potential collisions between satellites belonging to different operators, these services do not have access to the high-quality state information known to the operator of a satellite. Consequently, these screening services often have a high false alarm rate, with a detrimental impact on the trust placed on their alerts by satellite operators \citep{hiles_alexander_herzer_mckissock_mitchell_sapinoso_2021}. Miscommunications between operators have resulted in numerous near-misses between satellites in orbit  \citep{fouse_2021,grupen2021fairness}.  Motivated by these observations, we use our model to assess the value of sharing orbits and maneuver information among satellite operators.
\section{Related Work}
Collision avoidance of spacecraft has traditionally relied on optimal control approaches. For example, \cite{bombardelli_2015} relied on the linear relation between the applied thrust and an object's relative motion in the  collision plane to try to maximize the collision miss distance. Building upon \cite{bombardelli_2013}'s analytical formulation,  \cite{salemme_2020} designed low-thrust propulsion collision avoidance maneuvers using a finite-burn arc through an indirect optimal control model. \cite{vittori_2022} focused on computational efficiency and used an indirect trajectory optimization technique for on-board low-thrust collision avoidance maneuver design. The overarching assumption in these aforementioned works is that there is sufficient information about objects in the environment in order to efficiently maneuver and avoid collisions. By contrast, our work studies situations in which complete information may not be readily available. 

 In practice, information sharing is very limited in the space domain. The Federal Communications Commission and Department of Commerce solicited feedback from operators about space traffic management data sharing \citep{rendleman_mountin_2015,weeden_2019}; in response, commercial operators expressed a desire to limit the exchange of proprietary information that could give their competitors insight into the capabilities, health, and life of their satellites. It was also suggested that some operators may not have high-quality data to share. The above observations motivate the development of methods that can accommodate scenarios with limited information. 
 
 Recent work has used reinforcement learning for spacecraft trajectory optimization, guidance, and control \citep{izzo_martens_pan_2019, sullivan_bosanac_2020, cheng_wang_jiang_2019, hovell_ulrich_2020}. In contrast to these works, we focus on multi-agent coordination among satellites (or their operators) to improve the safety and efficiency of space traffic operations. While there has been much recent work on MARL, including for navigation and collision avoidance problems (\cite{MADDPG,MAPPO}; see \cite{informarl_icml} for more references), it has seen limited use in space domain applications \citep{peng_siew_sensor_tasking}. To the best of our knowledge, our work represents one of the first efforts to use MARL for satellite navigation and collision avoidance.  
    
\section{Methodology}

\begin{figure}
\hspace{-0.70cm}
    \includegraphics[width=15cm]{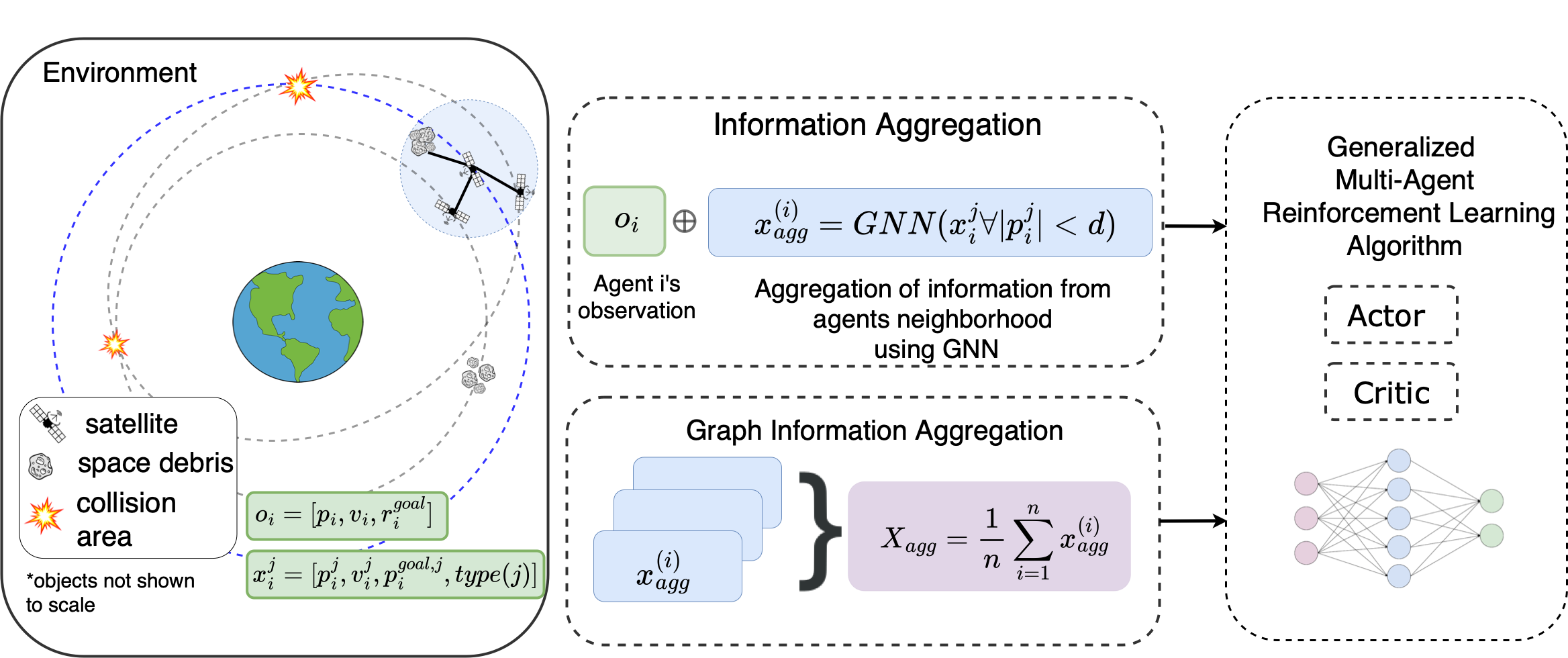}
    \caption{Schematic depiction of InforMARL, the approach that we adapt to the space environment in this paper. The meaning of each icon is shown in the legend  on the left of the environment box. Each agent or satellite $i$ receives observations about other satellites and space debris within its sensing radius, denoted by the blue shaded circle. This information is passed into a GNN, which learns compact representations of the relationships between objects. The output of the GNN is denoted $x^{(i)}_\mathrm{agg}$. (ii) Information aggregation: Each agent's state and observations ($o_i$) from a neighborhood $d$ are aggregated to obtain $X^{(i)}_\mathrm{agg}$. (iii) Graph information aggregation: The vectors of all the agents are averaged to get $X_\mathrm{agg}$. (iv) The concatenated vector [$o_i$, $X^{(i)}_\mathrm{agg}$] is input to an RL algorithm to obtain a recommended action. Figure adapted from \cite{informarl_icml}.
        \label{fig:mesh1}}
\end{figure}

Figure \ref{fig:mesh1} provides an overview of InforMARL \citep{informarl_icml}, the approach that we adapt for space traffic management applications. We make two main modifications to the original InforMARL method. First, we use a new space environment, described in Section \ref{environment}, that reflects the relative dynamics of satellites in orbit. Second, we modify the goal-sharing and goal-setting properties in the information aggregation and graph information aggregation modules of Figure \ref{fig:mesh1}, as explained in Section \ref{section:rep}.

\subsection{Environments}
\label{environment}

    In this work, we consider two different environments: (1) A ground environment in which the agents' dynamics are governed by a double integrator physics model \citep{double_integrator_model}, and (2) a space environment in which the relative motion between two agents (satellites) follows the Clohessy-Wiltshire equations \citep{vallado_mcclain_2007}. We also consider a modified space environment that accounts for perturbations in the dynamics caused by the oblateness of the Earth.

\subsubsection{Ground environment: Double-integrator model}
For the ground environment in our transfer learning experiments, we rely on a double integrator physics model to simulate the motion of agents ``on the ground" \citep{double_integrator_model}. Our environment is a modification of the one used in \cite{MAPE}. It corresponds to a 2D space in which agents move based on the following dynamics: 
\begin{align}
    \ddot{x} &= -\frac{\gamma}{m}\dot{x} + \frac{f_x}{m} \\
    \ddot{y} &= -\frac{\gamma}{m}\dot{y} + \frac{f_y}{m}
\end{align}

In the above, $f_x(t)$ and $f_y(t)$ represent the $x$ and $y$ components of the total force on an agent at time $t$. The total force is a sum of the control action from the RL algorithm (see point (iv) in Figure \ref{fig:mesh1}) and any collision forces experienced by the agent. $m$ is the mass of each entity, and $\gamma$ is a damping coefficient. In our simulations of the ground environment, each agent starts from a random, stationary position. The environment size is a hyperparameter, defining the area available for an agent to move in.

\subsubsection{Space environment: Clohessy-Wiltshire equations}
\begin{figure}[h!]
\centering
\begin{minipage}{.48\textwidth}
  \centering
  \includegraphics[width=0.9\linewidth]{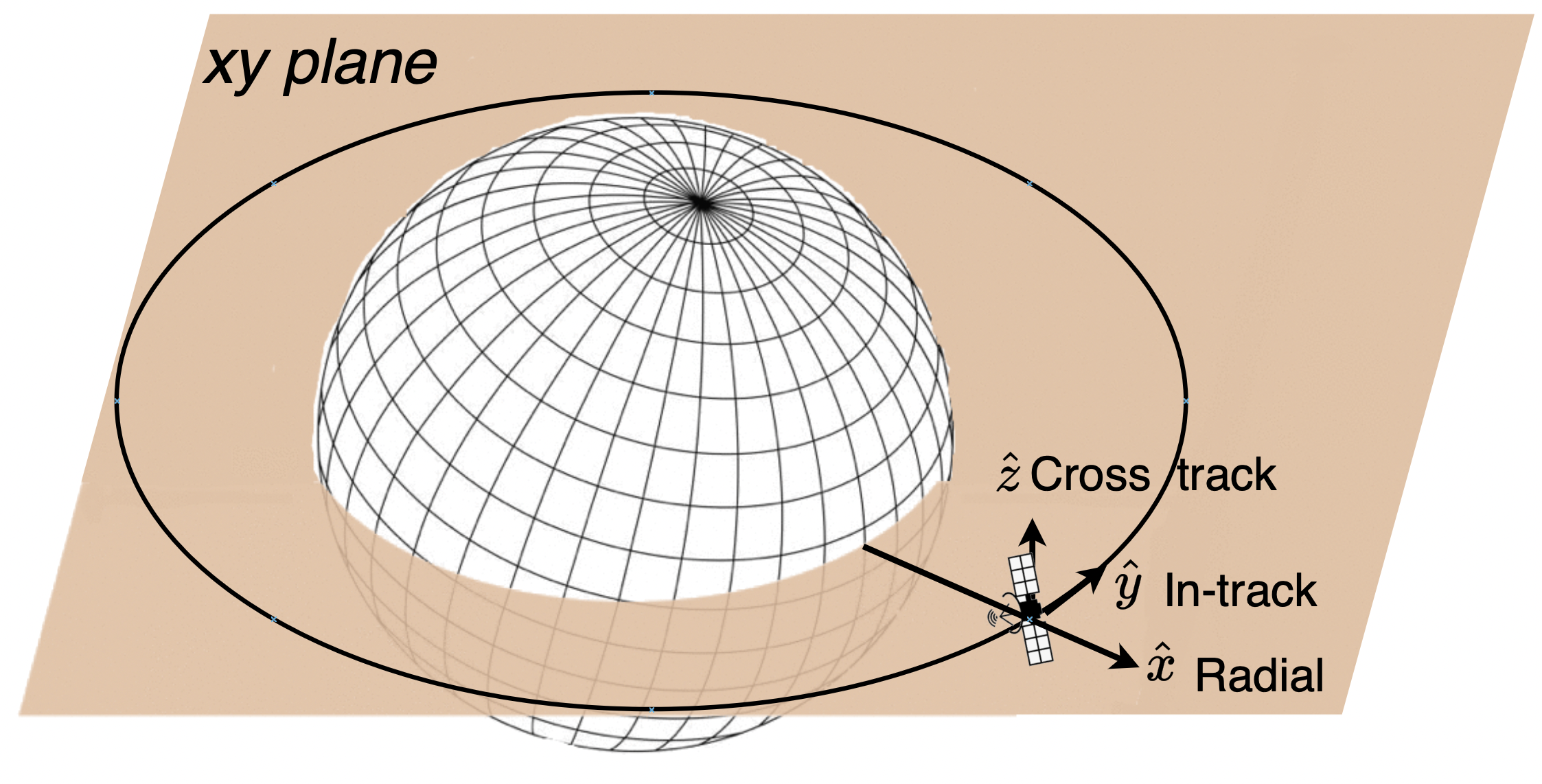}
  \captionof{figure}{The Radial In-track Cross-track (RIC) coordinate frame, centered on the target satellite. The radial direction is defined along the satellite's position vector, and the cross-track direction along its angular momentum vector. The in-track direction completes the triad. 
  }
  \label{hsw}
\end{minipage}%
\hfill
\begin{minipage}{.48\textwidth}
  \centering
  \includegraphics[width=0.75\linewidth]{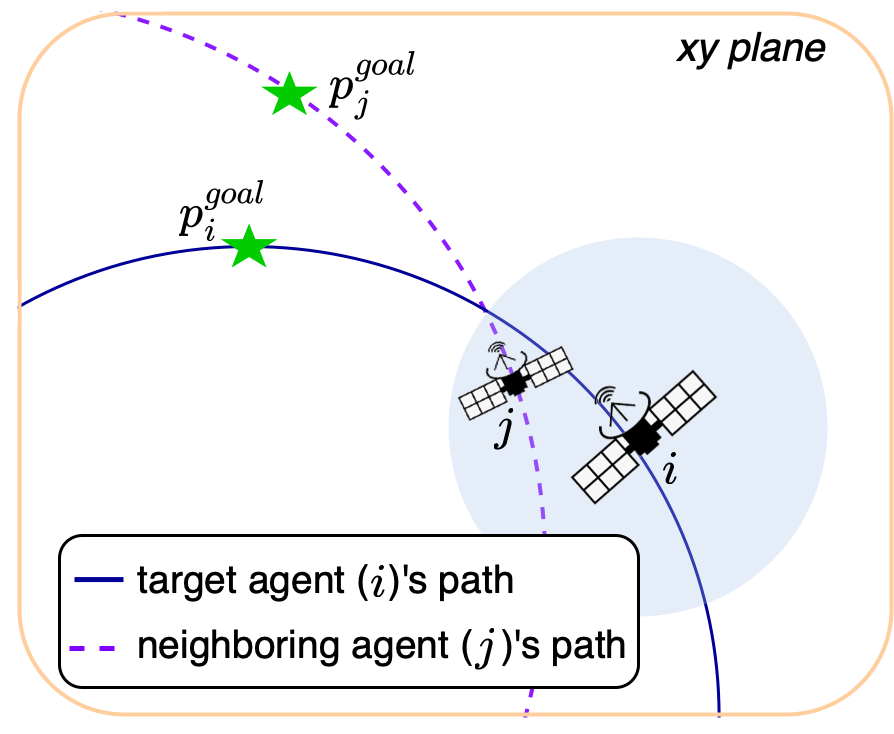}
  \vspace{-0.1in}
  \captionof{figure}{The $xy$ plane highlighted in Figure \ref{hsw}. The shaded circle represents the sensing region of satellite $i$. Each satellite has its own goal, denoted by  $p_{i}^{goal}$ and $p_{j}^{goal}$.}
  \label{man}
\end{minipage}
\end{figure}

For two objects in orbit, we consider their \emph{relative} motion to be governed by Clohessy-Wiltshire Equations \citep{vallado_mcclain_2007}, as follows:
    \begin{align}
    \ddot{x} &= 3\omega_n^2x+2\omega_n\dot{y} \label{xdotchw} \\
    \ddot{y} &= -2\omega_n\dot{x} \label{ydotchw} \\
    \ddot{z} &= -\omega_n^2z \label{zdotchw}
    \end{align}
The above equations consider a localized coordinate system centered around one of the satellites, referred to as the \emph{target}). The target satellite is assumed to have a circular orbit with an orbital rate of $\omega_n$. The coordinates are defined such that $x$ is measured radially outward from the target, $y$ is along the orbit track of the target body, and $z$ is along the angular momentum, as shown in Figure \ref{hsw}.

In general, spacecraft and debris have elliptical (including possibly circular) orbits. For example, suppose the target satellite in Figure \ref{man} has an orbit depicted by the solid line. The orbit of a nearby satellite on the same plane is shown by the dashed line.   Suppose the target satellite maneuvers to shift from its original orbit to a new transfer orbit, depicted by the dotted line. Since the coordinate frame is defined relative to the original frame, we can still track the relative motion of both satellites.

For the perturbed model, the dynamic equations are modified to include gravitational disturbances (or perturbations) due to Earth's oblateness: 
\begin{align}
    \ddot{x} &= (5c^2-2)\omega_n^2x+2\omega_nc\dot{y} \label{J2-x}\\
    \ddot{y} &= -2\omega_ncx \label{J2-y}\\
    \ddot{z} &= (2-3c^2)\omega_n^2z \label{J2-z}
\end{align}
In the above, $c$ is a parameter that reflects the change in orbital rate experienced by the satellite due to these perturbations. The value of $c$ is given by:
\begin{align}
   c &= \sqrt{1+s}, \text{ where } s = \frac{3J_2R_e^2}{8r^2}(1+3\cos(2\phi)). 
     \label{c}
\end{align}

In Equation \ref{c}, $J_2$ is a coefficient representing the gravitational effect of a body's oblateness, $R_e$ represents the Earth's radius, $r$ represents the  radius of the path of the target satellite, and the inclination, $\phi$, is the angle of the orbit relative to Earth's equator. An angle of $\phi=0^\circ$ represents an equatorial orbit, whereas $\phi=90^\circ$  represents a polar orbit. We note that substituting $c=1$ (i.e., $\phi=54^\circ$) in Equations \eqref{J2-x}-\eqref{J2-z} results in the Clohessy-Wiltshire equations (\eqref{xdotchw}-\eqref{zdotchw}).

A full derivation of the perturbed equations can be found in \cite{cw_j2}. While there exist additional environmental perturbations, such as solar radiative pressure or three-body effects, their impacts are magnitudes smaller \citep{walter_2018}. Consequently, these other perturbations only affect satellite dynamics over time periods that are much longer than the episode lengths considered in this paper, which are of the order of a few hours.

This work focuses on in-plane maneuvers, which means all satellites and debris involved are assumed to lie on the same $xy$ orbital plane, as shown in Figure \ref{hsw}. Similar to the ground environment, the maneuvers are determined by the control action (a force in the $xy$ plane) recommended by the RL algorithm. 
In both the original Clohessy-Wiltshire equations and the perturbed model, $z$-dynamics is decoupled from those in the $x$- and $y$-directions. While not a topic of investigation in this paper, we believe that, in principle, the same techniques could be applied to the cross-track dimension.

\subsubsection{Comparison of ground and space environments\label{compare}}

We explore the potential of transfer learning for this application because the space environment has complicated dynamics that are much more challenging for the algorithm to learn. From equations \ref{xdotchw} and \ref{ydotchw}, we see that the $x$ and $y$ variables in the Clohessy-Wiltshire equations are coupled. By contrast, the ground equations are independent of one another. 

In both cases, we adopt a simplistic reward function similar to the one used in multi-agent particle environment  \citep{MAPE}. We assume that at time $t$, each agent $i$ gets a reward: $r_t^{(i)} = r_{\mathrm{dist},t}^{(i)} + r_{\mathrm{coll},t}^{(i)} + r_{\mathrm{goal},t}^{(i)}$, where $r_{\mathrm{dist}, t}^{(i)}$ is the negative of the Euclidean distance to the goal, $r_{\mathrm{coll},t}^{(i)}=-5$ if it collides with any other entity and zero otherwise, and $r_{\mathrm{goal},t}^{(i)}=+5$ if the agent has reached the goal and zero otherwise. The joint reward function is defined as the sum of the individual agent rewards, which encourages cooperation among all agents. It is worth noting that reward function can be further refined, especially in terms of how collisions are penalized. Future research will include imposing larger penalties for collisions, as well as the use of control barrier functions for providing safety guarantees \citep{cbf}.

It should be noted that the scale of distances in the ground environment is in meters (and speeds in m/s), whereas the scale of the space environment is in kilometers (and km/s, respectively). Despite the orders-of-magnitude differences in the units, the numerical values are quite similar in the ground and space environments. Although the control action is the same  (1 N) for both environments, the dynamics in the space environment are  much more sensitive to external control actions \citep{magman}. We investigated additional scaling mechanisms between the two environments, but found that the sample complexity of InforMARL was sufficient to achieve successful transfer learning.

\subsection{Graph representation} 
    \label{section:rep}
    
    We use an agent-entity graph as used by \cite{EMP, informarl_icml} to represent the interactions between different satellites and debris in the environment. For each agent $i$, we define an agent-entity graph at each time-step $t$. This graph is made up of nodes and edges, $g^{(i)}_t \in \mathcal{G}: (\mathcal{V}, \mathcal{E})$, where each node $v\in\mathcal{V}$ is an entity in the environment. Edges, $e\in\mathcal{E}$, are defined when an agent and an entity are within a sensing radius $d$ of one another. For agent-agent interactions, the edges are bidirectional, meaning that the messages are passed back and forth. For agent-non-agent interactions, the edges are unidirectional, meaning that messages are only passed from the non-agent entity to the agent. Translating this to the space environment, a bi-directional edge means that there is open communication between both satellites, while a unidirectional edge means that the satellite has received information about the location of a nearby piece of debris.  Each agent $i$'s local observation $o^{(i)}$ consists of its position and velocity in a global frame of reference and the relative position of the agent's goal with respect to its position.

    As depicted in the environment block in Figure \ref{fig:mesh1}, each node $j$ in the graph $g^{(i)}$ has node features $x_j=[p^j_i, v^j_i, p^{\mathrm{goal},j}_i, \texttt{entity\_type}(j)]$ where $p^j_i, v^j_i, p^{\mathrm{goal},j}_i$ are the \emph{relative} position, velocity, and position of the goal of the entity at node $j$ with respect to agent $i$, respectively. If node $j$ is an obstacle or a goal, then it is set to be equivalent to the position  $p^{\mathrm{goal},j}_i \equiv p^j_i$. The variable \texttt{entity\_type}$(j)\in\{ \texttt{agent}, \texttt{obstacle}, \texttt{goal}\}$ determines the type of entity at node $j$. In this context, the $\texttt{goal}$ entity type is used to indicate each satellite's intentions to maneuver.

    To determine the value of information sharing, we rely on two graph variants. The first graph variant is for the case when satellites share their goals. For this graph type, the agents share both their goal $p^{\mathrm{goal},j}_i$ and their state information $p^j_i, v^j_i$. The second graph variant is for the case when satellites are hiding their goals. For this graph type, each agent node $j$ on the graph is transformed into: $x_j=[p^j_i, v^j_i, \texttt{entity\_type}(j)]$.

\section{Numerical Experiments}
In our numerical experiments, we first demonstrate the scalability of the approach using the space environment for both training and testing, while varying the number of agents. Next, we consider the effectiveness of transferring a model trained in the ground environment to the space one. Finally, we use our transfer learning-based model to evaluate the benefits of satellite operators sharing maneuver information for the purposes of space traffic management.

 We first demonstrate the scalability of the approach using the space environment for both training and testing, while varying the number of agents. Next, we consider the effectiveness of transferring a model trained in the ground environment to the space one, both in unperturbed and perturbed cases. We then use our model to evaluate the benefits of satellite operators sharing goal information for the purposes of space traffic management.
    
 Agents start at random locations at the beginning of each episode; the corresponding goals are also randomly distributed. Static obstacles are placed randomly in the environment in each episode. The environment is a 2 km $\times$ 2 km area, and the episode length is approximately one hour. Each scenario is initialized with three pieces of debris, which remain at the same relative locations throughout the simulation. The episode continues even after a collision. However, a collision force is exerted on any colliding agent that affects its subsequent dynamics. The overarching objective is for every agent to reach its corresponding goal without colliding with any other entity.  Since the episode continues even after a collision, it is possible for there to be multiple collisions in the same episode. 
 
 We calculate the following metrics: 1) The total rewards obtained by the agents during an episode using the reward function defined in Section \ref{compare}. A higher value corresponds to better performance. 2) The fraction of an episode taken on average by agents to get to their goals, denoted $T$ (lower is better). $T$ is set to 1 if an agent does not reach its goal. 3) Percent of episodes in which all agents get to their goals, denoted $S$\% (higher is better). 4) The total number of collisions 
 those agents had in an episode, denoted \# col. The lower this metric, the better the performance of the algorithm. Although having fewer collisions is better, some learned  policies do not significantly move the agents from their initial orbit, and hence do not get to the goal but have a lower collision rate. In short, the collision rate and the success rate should be considered in conjunction with each other. 

It is important to note that the size of the environment considered (4 sq km) along with the number of objects in it (between 6-13) result in unrealistically dense scenarios. The purpose of these experiments is to evaluate the scalability of the methods and the general trends in performance, and not to determine values (e.g., of the collision rates) that are  representative of real-world operations.

\begin{table}[h!]
        \centering
        \footnotesize
           \begin{tabular}{|l|c||c|c|c|} \hline 
            \multicolumn{2}{|c||}{\diagbox{\bf Train}{\bf Test}}& $\boldsymbol{m}\mathbf{=3}$  & $\boldsymbol{m}\mathbf{=5}$ &  $\boldsymbol{m}\mathbf{=10}$\\ \hline\hline 
                \multirow{4}{*}{$\boldsymbol{n}\mathbf{=3}$} 
                & Reward/$m$ & 61.57 & 60.21 &57.78 \\ \cline{2-5}
                & $T$ & 0.44 & 0.44 & 0.43 \\ \cline{2-5}
                & (\# col)/$m$ & 0.36 & 0.77 &1.41  \\ \cline{2-5}
                & $S$\% & 98 & 94 & 96 \\ \hline\hline 
                \multirow{4}{*}{$\boldsymbol{n}\mathbf{=5}$} 
                & Reward/$m$ & 60.52 & 60.52 & 57.07 \\ \cline{2-5}
                & $T$ & 0.44 & 0.44 & 0.44 \\ \cline{2-5}
                & (\# col)/$m$ & 0.78 & 1.28 & 1.41 \\ \cline{2-5}
                & $S$\% & 98 & 98 & 91 \\ \hline
            \end{tabular}  
            \captionof{table}{Performance metrics obtained by training InforMARL on a space environment with $n$ satellites and testing it on one with $m$ satellites: (a) Total reward obtained in an episode per agent, Reward$/m$. (b) Fraction of episode taken on average by agents to reach their goal, $T$ (lower is better). (c) Average number of collisions per agent in an episode, \#col$/m$ (lower is better). (d) Success rate, $S$\%: percentage of episodes in which all agents are able to get to their goals (higher is better).}
            \label{scale_Table}
    \end{table}
 
\subsection{Scalability}
      
    \label{sec:scalability}
   The first experiment, shown in Table \ref{scale_Table}, demonstrates the scalability of our algorithm when trained on $n$ agents and tested on $m$ agents. The number of obstacles is held constant to be 3 throughout both all training and evaluation. When evaluating the connectivity throughout the experiments, we found that each agent maintained a connection with at least one entity in the environment more than $90$\% of the time, meaning that the vast majority of the experiment involves information-sharing between entities. We find that in all scenarios considered, our approach can control the satellites to reach their goals within approximately 44\% of the episode length. As expected, the number of collisions per agent increases when there are more satellites (i.e., the environment has become more dense). As mentioned in Section \ref{compare}, these values can be improved further by modifying the reward function or by using control barrier functions. 
        A key finding is that the reward per agent remains approximately the same even when the model is trained with $n<m$. Our approach also has robust sample complexity, with a high success rate for unseen scenarios. As the number of objects in space is expected to dramatically increase in the coming years, the scalability and sample efficiency of our technique make it a promising approach to space traffic management.

\subsection{Transfer learning}
 \begin{figure}[ht!]
  \centering
        \includegraphics[width=\linewidth]{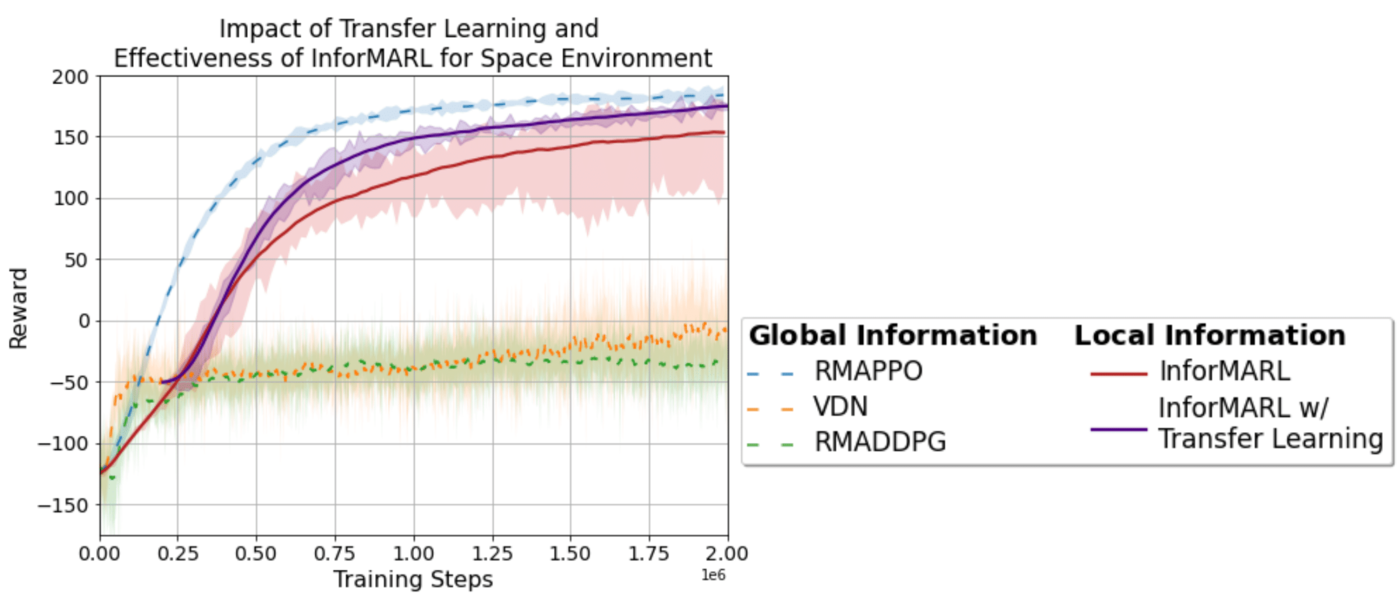}
          \caption{InforMARL with transfer learning learns a similar reward to RMAPPO despite having access to less information. The transfer learning model also outperforms the baseline InforMARL model that was trained only on a space environment.}
      \label{baseperformance}
    \end{figure}
        Figure \ref{baseperformance} compares the effectiveness of transfer learning (denoted by InforMARL w/ Transfer Learning) relative to other MARL baseline algorithms. The transfer learning model was initialized using weights trained on a ground environment, and then trained in the space environment. Consequently, in Figure \ref{baseperformance}, we offset the plot of InforMARL with transfer learning by 200,000 steps (the number of steps used to train the ground-based model). We trained each algorithm 5 times on 5 separate seeds. The shaded area envelops one standard deviation of the runs.  
        
        Three of the baseline methods (RMAPPO \citep{MAPPO}, VDN \citep{VDN} and RMADDPG \citep{MADDPG}) use global information, i.e., every satellite in the environment would be required to share their information publicly. We also explored the performance of QMIX \citep{QMIX} and MATD3 \citep{MATD3}, but the performance was similarly poor. While the assumption of global information sharing can help determine a performance bound, it is not realistic in practice. By contrast, InforMARL (both with and without transfer learning) uses only local information. Over the training period, InforMARL with transfer learning reaches a similar reward to RMAPPO \citep{MAPPO}, despite needing significantly less information. This finding indicates that the quality of the information, rather than its quantity, is an important driver of performance. Furthermore, we see in Figure \ref{baseperformance} that InforMARL with transfer learning outperforms the InforMARL model that was trained from scratch on the satellite environment. These results indicate that ground-based models can be used to accelerate training for space traffic applications. Initial investigations suggest that the poor performance on some instances without transfer learning are because of allocation to weaker computational nodes. InforMARL with transfer learning appears to be less impacted by the specific computational nodes assigned for training. A possible reason is that the initialization with the ground model helps avoid some of the less efficient learning paths, resulting in more consistent training performance. We plan to study this further in future work.

        Based on the above results, we believe that the benefits of using InforMARL with transfer learning for such space applications are mainly two-fold: 
        (1) The simplicity of the ground dynamics and the availability of more established ground simulation environments make transfer learning from ground to space an attractive approach; and (2) InforMARL with transfer learning appears to be less susceptible to the performance of the specific computational nodes that are assigned for training.

\subsection{Sensitivity To Perturbations.}

    Table \ref{Sensi} demonstrates the performance of InforMARL with transfer learning in the perturbed satellite environment for 3 agents. Once again, the transfer learning model was initialized using weights trained on a ground environment, and then trained in the space environment. We find that even in perturbed environments, the sample complexity of InforMARL allows the algorithm to learn effectively.

     In the perturbed model with dynamic equations given by \eqref{J2-x}-\eqref{J2-z}, an inclination of 54\textdegree corresponds to the unperturbed Clohessy-Wiltshire equations. When comparing the additional inclination values on the table to 54\textdegree, the transfer learning model still performed well, at a comparable level to the model without perturbations ( Table \ref{Sensi}). 
    \begin{table}[h!]
    \centering
    \footnotesize
      \begin{tabular}{|c||r|r|r|r|r|r|r|r|r|} \hline 
        {\bf Inclination} ($\boldsymbol{\phi}$) & {\bf 0\textdegree} &  {\bf 28\textdegree} & {\bf 45\textdegree}& {\bf 54\textdegree} &  {\bf 63\textdegree}& {\bf 72\textdegree}& {\bf 81\textdegree} &  {\bf 90\textdegree} \\ \hline \hline
        {\bf Average reward} & 170.10 & 166.33 & 167.07& 169.73& 168.14& 171.78 & 174.29 & 169.76 \\ \hline
        {\bf Standard deviation} & $\pm$ 6.32 & $\pm$ 6.18 &$\pm 6.09$ &$\pm$ 3.07  & $\pm$ 8.85 & $\pm$ 5.45 & $\pm$ 3.02 & $\pm$ 6.10 \\\hline   
        \end{tabular}
        \caption{Sensitivity of InforMARL with transfer learning to different perturbations (orbit inclinations), for a three-agent system. The average total reward achieved at the end of training and its standard deviation are presented, based on 5 runs with 5 different random seeds.}
        \label{Sensi}
\end{table}

\subsection{Value of sharing goals}

\begin{figure}[h!]
\begin{minipage}{0.35\textwidth}
  \centering
  \includegraphics[width=\linewidth]{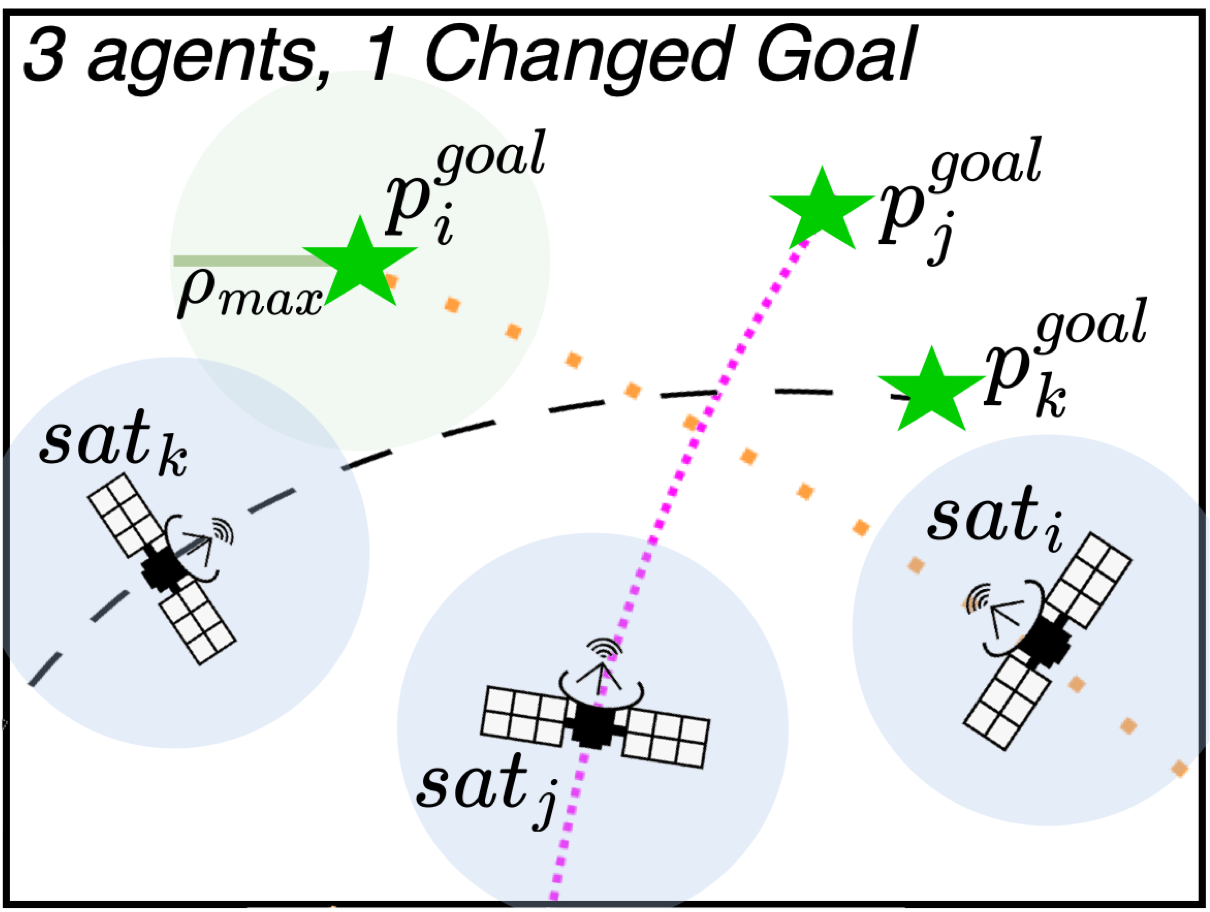}
  \captionof{figure}{For one of the three agents, the goal is randomly resets  midway through the episode to be within a distance $\rho_{\max}$ from $p_{i}^{goal}$. \label{GoalSharing}}
  \label{fig:test1}
\end{minipage}%
\hspace{4mm}
\begin{minipage}{0.61\textwidth}
  \centering
  \includegraphics[width=\linewidth]{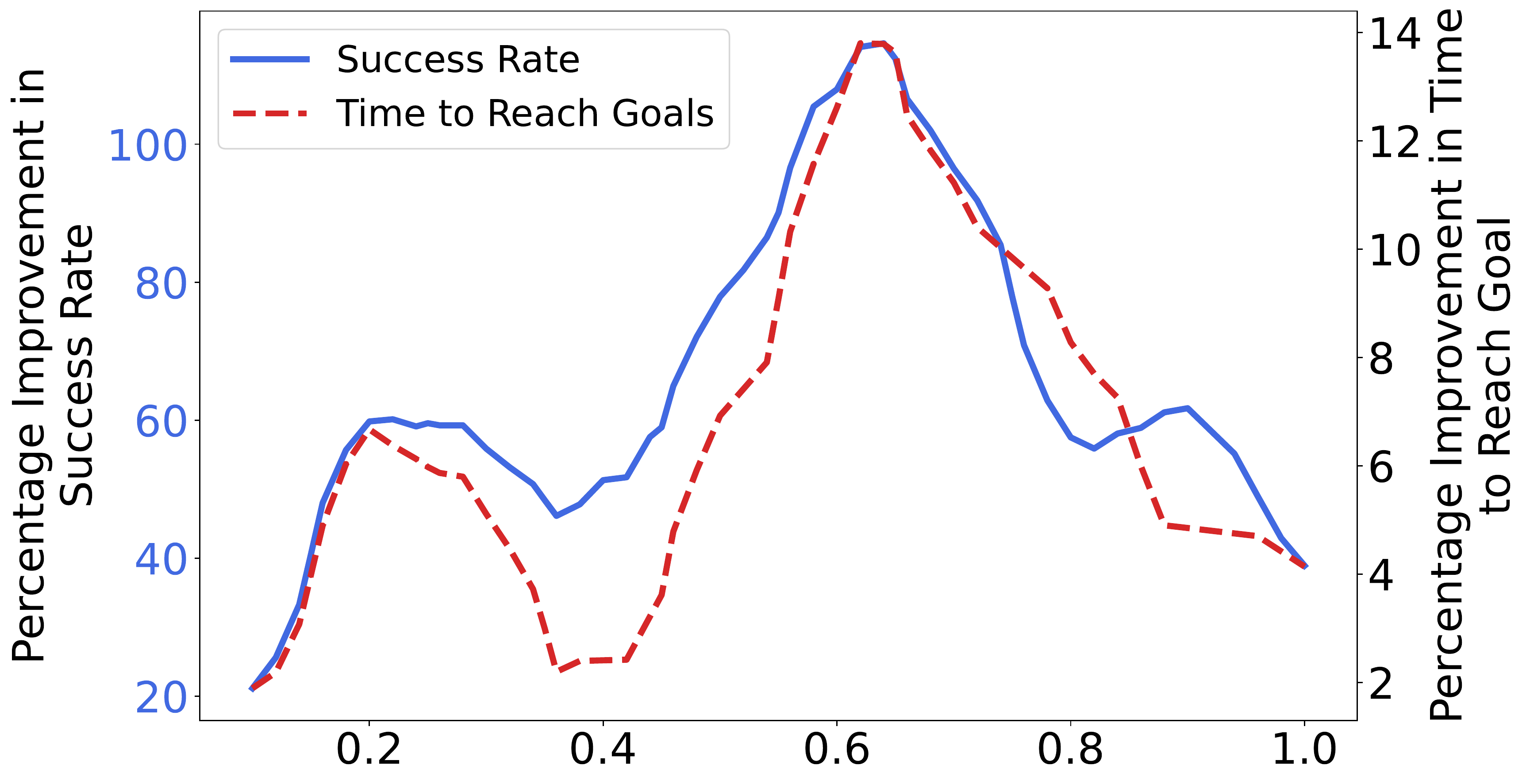}
  \captionof{figure}{Percentage improvement achieved through goal-sharing for (1) the success rates, $S$ (in blue; left-axis), and (2) the fraction of episode (or time) taken on average by agents to reach their goal, $T$ (in red; right-axis). Moving averages over 0.2 km increases in $\rho_{\max}$ are shown.}
  \label{fig:Distribution}
\end{minipage}
\end{figure}

We use our method to evaluate the value of satellites sharing their goals (i.e., sharing their orbits, and any associated changes or maneuvers) with each other. In our experiments, a randomly chosen satellite changes its goal midway through the episode. The new goal is set to a uniformly-random location that is within a distance $\rho_{\max}$ of the original goal, as illustrated in Figure \ref{GoalSharing}. 

We progressively increase $\rho_{\max}$ in steps of 20 m from zero to 1 km (which considering the 2-km$\times$2-km environment size is approximately equivalent to resetting  the goal randomly somewhere within the environment). We generate 100 instances for each value of $\rho_{\max}$, with random initializations of satellite locations and their goals. For each such instance, we use InforMARL trained on the space environment to evaluate the cases when the satellites share their goals with each other, and when they do not. We consider two performance measures from Section \ref{sec:scalability}, the success rate ($S$) and the average time to reach goal ($T$). We evaluate the performance improvement from goal-sharing as follows: For the success rates, we calculate $\frac{S_{\mathrm{goal\_sharing}} - S_{\mathrm{no\_goal\_sharing}}}{S_{\mathrm{no\_goal\_sharing}}}$ as a percentage. The improvement in the average time to reach the goal is calculated similarly, accounting for the fact that an improvement corresponds to a decrease in $T$. 

Figure \ref{fig:Distribution}  demonstrates the performance improvement (relative to the performance without goal sharing) that is achieved through goal sharing, as the maximum goal reset distance increases. We note that positive values in Figure \ref{fig:Distribution} indicate that the success rates \emph{increase} with goal-sharing and the times taken by agents to reach their goals \emph{decrease}, illustrating the benefits of goal-sharing for all values of $\rho_{\max}$ considered. The median number of collisions was found to be the same irrespective of goal-sharing, so we do not discuss this performance metric further. 

For small values of $\rho_{\max}$ (e.g., $\rho_{\max} < 0.3$ km in Figure \ref{fig:Distribution}), the approximate location of the new goal can be inferred from the current goals/orbits, so we see relatively modest performance improvement in success rates from goal-sharing. As $\rho_{\max}$ increased further, we see significant improvements (reaching a more than 100\% increase) in the success rates from goal-sharing. Interestingly, we see that for higher values of $\rho_{\max}$ (e.g., $\rho_{\max} > 0.65$ km in Figure \ref{fig:Distribution}), the benefits of goal-sharing begin to decrease. We hypothesize that this behavior is related to relationship between the aleatory uncertainty of satellite trajectories and collision risk \citep{Balch2019}. Let us consider the example described by \cite{Balch2019}: Suppose two satellites are known with certainty to be on a collision course (i.e., there is no epistemic uncertainty). If we knew that one of the operators could apply an arbitrarily large maneuver in a random, unknown direction, then the probability of a collision post-maneuver decreases as the magnitude of the possible maneuver increases. Analogously in our situation, this would mean that for large enough values of $\rho_{\max}$, the success rates start increasing even without goal-sharing; consequently, the percentage improvement in these rates obtained from goal sharing starts decreasing as seen in Figure \ref{fig:Distribution}.

In summary, we find that goal-sharing generally improves success rates and decreases times to goal, demonstrating the potential benefits of such information-sharing among satellite operators.

\section{Conclusions and Future Work}
\label{section:conclusion_future_work}
We demonstrated that transfer learning from a ground-based environment to a space-based one can improve both sample complexity and performance, despite the significant differences in the underlying dynamics that govern the agents in the two environments. We also found that InforMARL, our GNN-based approach, is scalable in a space-based environment, satisfying a critical need for space traffic management as the skies become more dense with satellites and debris. Our initial investigations showed that when goal changes are bounded, our method can leverage goal sharing among satellite operators to improve the safety and efficiency of space traffic operations. Future work will include developing a more realistic space traffic simulation environment, accounting for communication delays and losses, adding mechanisms to provide safety guarantees, and designing incentives for information-sharing.

\section*{Acknowledgements}
    The authors would like to thank the MIT SuperCloud \citep{supercloud} and the Lincoln Laboratory Supercomputing Center for providing high performance computing resources that have contributed to the research results reported within this paper. Additionally, we would like to thank Jasmine Aloor and Victor Qin for their help running some of the simulations. The NASA University Leadership Initiative (grant \#80NSSC20M0163) provided funds to assist the authors with their research, but this article solely reflects the opinions and conclusions of its authors and not any NASA entity. This research was sponsored in part by the United States AFRL and the United States Air Force Artificial Intelligence Accelerator and was accomplished under Cooperative Agreement Number FA8750-19-2-1000. The views and conclusions contained in this document are those of the authors and should not be interpreted as representing the official policies, either expressed or implied, of the United States Air Force or the U.S. Government. The U.S. Government is authorized to reproduce and distribute reprints for Government purposes notwithstanding any copyright notion herein. Sydney Dolan was supported by the National Science Foundation Graduate Research Fellowship under Grant No. 1650114.

\clearpage
{\small
\bibliographystyle{unsrt}
\bibliography{references} 
}


\newpage

\appendix
\section{Baseline Implementation Sources}
\label{appendix:implementation}
We modified the codebases from the official implementations for the MAPPO baselines. We also adapted the codebase for MADDPG and  VDN for our experiments which was benchmarked on other standard environments. The links for all these implementations are listed below. 

\begin{itemize}
    \item MAPPO: \href{https://github.com/marlbenchmark/on-policy}{https://github.com/marlbenchmark/on-policy}
    \item MADDPG, VDN:
    \href{https://github.com/marlbenchmark/off-policy}{https://github.com/marlbenchmark/off-policy}

\end{itemize}

\section{Hyperparameters}
\label{appendix:hyperparams}

We performed a hyperparameter search for these algorithms by varying the learning-rates, network size and a few algorithm specific parameters. We observed that the hyperparameters used in the original implementation gave the best performance, and used those same values for our experiments as well.

Table \ref{table:informarl-hp} lists the hyperparameters specific to the InforMARL implementation. 
Here, ``entity embedding layer dim" and ``entity hidden dim" refer to the embedding layer input and output dimension respectively which is used to process the \texttt{entity-type} variable in the graph. ``add self loop" refers to whether a self-loop should be added while constructing the agent-entity graph. ``gnn layer hidden dim" is the output dimension of each layer in the graph transformer. ``num gnn heads" and ``num gnn layers" are the number of heads in the attention layer and number of attention layers used in the graph transformer. ``gnn activation" is the activation function used after each layer in the GNN module.

\begin{table*}[h]
  \centering
  \begin{tabular}{lc}
    \toprule
    Hyperparameters     & Value \\
    \midrule
    entity embedding layer dim     &  3 \\
    entity hidden dim     &  16 \\
    num embedding layer     &  1 \\
    add self loop     &  False \\
    gnn layer hidden dim     &  16 \\
    num gnn heads     &  3 \\
    num gnn layers &  2\\
    gnn activation     &  ReLU \\
    \bottomrule
  \end{tabular}
  \caption{Hyperparameters used in InforMARL}
  \label{table:informarl-hp}
\end{table*}

Tables \ref{table:mappo-hp}, \ref{table:offpolicy-hp}, \ref{table:all-hp} list the hyperparameters common for the InforMARL, MAPPO, MADDPG, and VDN implementations. For MADDPG, MAPPO, VDN and InforMARL, ``batch size” refers to the number of environment steps collected before updating the policy via gradient descent. “mini batch” refers to the number of mini-batches a batch of data is split into. “gain” refers to the weight initialization gain of the last network layer for the actor network. ``num envs" refers to the number of parallel roll out threads used to collect state-transition tuples.

\begin{table*}[h]
  \centering
  \begin{tabular}{lc}
    \toprule
    Common Hyperparameters     & Value \\
    \midrule
    recurrent data chunk length & 10 \\
    gradient clip norm     & 10.0 \\
    gae lambda     & 0.95 \\
    gamma     & 0.99 \\
    value loss     &  huber loss \\
    huber delta     &  10.0 \\
    batch size     &  num envs $\times$ buffer length $\times$ num agents\\
    mini batch size     &  batch size / mini-batch\\
    gain & 0.01 \\
    network initialization     &  Orthogonal \\
    optimizer     &  Adam \\
    optimizer epsilon     &  1e-5 \\
    weight decay     &  0 \\
    use reward normalization     &  True \\
    use feature normalization     &  True \\
    \bottomrule
  \end{tabular}
  \caption{Common Hyperparameters used in MAPPO and InforMARL}
  \label{table:mappo-hp}
\end{table*}

\begin{table*}[h]
  \centering
  \begin{tabular}{lc}
    \toprule
    Common Hyperparameters     & Value \\
    \midrule
    gradient clip norm & 10.0 \\
    random episodes &  5 \\
    epsilon &  1.0 $\rightarrow$ 0.05\\
    epsilon anneal time &  50000 timesteps\\
    train interval & 1 episode \\
    gamma & 0.99 \\
    critic loss & mse loss \\
    buffer size & 5000 episodes \\
    batch size & 32 episodes \\
    optimizer & Adam \\
    optimizer eps & 1e-5 \\
    weight decay & 0 \\
    network initialization & Orthogonal \\
    use reward normalization & True \\
    use feature normalization & True \\
    \bottomrule
  \end{tabular}
  \caption{Common Hyperparameters used in MADDPG, VDN}
  \label{table:offpolicy-hp}
\end{table*}

\begin{table*}[h]
  \centering
  \begin{tabular}{lc}
    \toprule
    Common Hyperparameters     & Value \\
    \midrule
    num envs     & 128 \\
    buffer length     &  25 \\
    num GRU layers     &  1 \\
    RNN hidden state dim     &  64 \\
    fc layer hidden dim     &  64 \\
    num fc     &  2 \\
    num fc after     &  1\\
    \bottomrule
  \end{tabular}
  \caption{Common Hyperparameters used in MAPPO, MADDPG, VDN and InforMARL}
  \label{table:all-hp}
\end{table*}

\section{Computational Environment} 

Our models were trained on a server with 40 Intel Xeon Gold 6248 @ 2.50 GHz processor cores and 2 NVIDIA Volta V100 graphics cards. Our code uses PyTorch \cite{pytorch} version 1.11.0, CUDA version 11.3,
and PyTorch Geometric Library \cite{PyG} version 2.0.4. 

\end{document}